\documentclass[twocolumn,aps,prd,nofootinbib,superscriptaddress,showkeys]{revtex4-1}
\usepackage[labelsep=quad,indention=10pt,caption=false]{subfig}
\usepackage[colorlinks=true,linkcolor=black,citecolor=blue,urlcolor=black]{hyperref}
\usepackage{graphicx}
\usepackage{longtable}
\usepackage{amstext,amsmath,amsfonts,amssymb,amsthm}
\usepackage{color}

\usepackage{verbatim}
\def\l{\left}
\def\r{\right}

\def\eq#1{{Eq.~(\ref{#1})}}

\def\fig#1{{Fig.~\ref{#1}}}

\def\be{\begin{equation}}
\def\ee{\end{equation}}
\def\bes{\begin{eqnarray}}
\def\ees{\end{eqnarray}}
\def\ba{\begin{align}}
\def\ea{\end{align}}
\def\bwt{\begin{widetext}}
\def\ewt{\end{widetext}}
\def\tbfk{\textbf{k}}

\def\aa{\alpha}
\def\kk{\kappa}
\def\aa{\alpha}
\def\tt{\tau}

\def\oo{\omega}

\def\bra{\langle}
\def\ket{\rangle}
\def\f{\frac}

\def\pp{\partial}
\def\nn{\nonumber}

\begin{document}

\title{Multifaceted Schwinger effect in de Sitter space}

\author{Ramkishor Sharma}
\email{rsharma@physics.du.ac.in}
\email{sharmaram.du@gmail.com}
\affiliation{Department of Physics \& Astrophysics, University of Delhi, New Delhi 110 007 India.}

\author{Suprit Singh}
\email{suprit.singh@unb.ca}
\email{supritsingh@gmail.com}
\affiliation{Department of Physics \& Astrophysics, University of Delhi, New Delhi 110 007 India.}
\affiliation{Department of Mathematics \& Statistics, University of New Brunswick, Fredericton E3B 5A3 Canada.}

\begin{abstract}
\noindent  We investigate particle production \`a la Schwinger mechanism in an expanding, flat de Sitter patch as is relevant for the inflationary epoch of our universe. Defining states and particle content in curved spacetime is certainly not a unique process. There being different prescriptions on how that can be done, we have used the Schr\"odinger formalism to define \emph{instantaneous} particle content of the state etc. This allows us to go past the adiabatic regime to which the effect has been restricted in the previous studies and bring out its \emph{multifaceted} nature in different settings. Each of these settings gives rise to contrasting features and behaviour as per the effect of electric field and expansion rate on the \emph{instantaneous} mean particle number. We also quantify the degree of classicality of the process during its evolution using a ``classicality parameter" constructed out of parameters of the Wigner function to obtain information about the quantum to classical transition in this case.  

\end{abstract}

\keywords{Particle creation, Electric Fields, Inflation, de Sitter space, Schr\"odinger quantization, quantum-to-classical transition.}

\maketitle

Strong electric fields can cause the ``quantum" vacuum to decay into charged pairs -- an effect in quantum field theory first predicted by Julian Schwinger \cite{Schwinger:1951a} and aptly known by his name. It is also a subject of hot pursuit for an experimental verification \cite{Dunne:2010}. Analogous to the effect of electromagnetic fields, we also have gravitational particle production in the study of quantum fields in curved spacetimes \cite{books, reviews, papers}. A particularly important and specific case of particle production due to time-dependent gravitational background is during inflation or in de Sitter space \cite{desitter}. There is a general consensus that the large scale structures and the anisotropies of the Cosmic Microwave Background have their origin in the early inflationary phase of the Universe \cite{lyth}. We also observe large scale magnetic fields in the universe with coherent lengths extending from few kpc to Mpcs and strength varying between $\mu G$ to $nG$ \cite{magobs}. Some recent observations also suggest the presence of these magnetic fields in voids \cite{Neronov}. The origin of these magnetic fields is still an open question. One of the widely accepted view \cite{magnetic} is that these were generated during inflation possibly by breaking conformal invariance and hence have a primordial origin. In light of this, gravitational and the electromagnetic fields coexisting during inflation will have a combined effect on the vacuum of any (test) quantum field propagating on the background. This forms precisely the setting for Schwinger effect in de Sitter space. 

The effect was considered in connection with neutralization of the cosmological constant through membrane creation  \cite{Brown:1988} and then for computing spontaneous nucleation rates \cite{Garriga:1994} in inflation. There have also been other investigations \cite{dsads} with a take on anti-de Sitter space as well. Some very recent works, however, have examined the effect in the form more relevant to the inflationary physics \cite{inflationary, Kobayashi:2014, inflationary1}. Our analysis has a similar perspective i.e., a connection with inflation but with a difference that expands the existing results significantly. This calls for some comments on the previous works and on the framework adopted in this article. 

To put things in perspective, first note that in a generic curved background and particularly cosmological spacetimes including the (quasi-) de Sitter phase of inflation, there is an explicit time-dependence in the metric and the time-like killing vectors are non-existent. We also do not have the luxury to switch off the background effects, that is, there is no flat spacetime limit asymptotically in time. As such, the usual prescription of defining the \emph{in} and \emph{out} vacuum states \emph{does not work} and the definition or notion of a vacuum is not unique. Nonetheless, one can still define and compute the time evolution of a quantum state without much ambiguity. However, determining the particle content of this state at any given time during the course of its evolution can be done in different ways open to many interpretations. At best, we can infer it using different constructs that probe the physics in time-dependent background and develop an intuitive feel for the various phenomenon. 

\begin{figure*}[th!]
\centering
\includegraphics[width=0.48\textwidth]{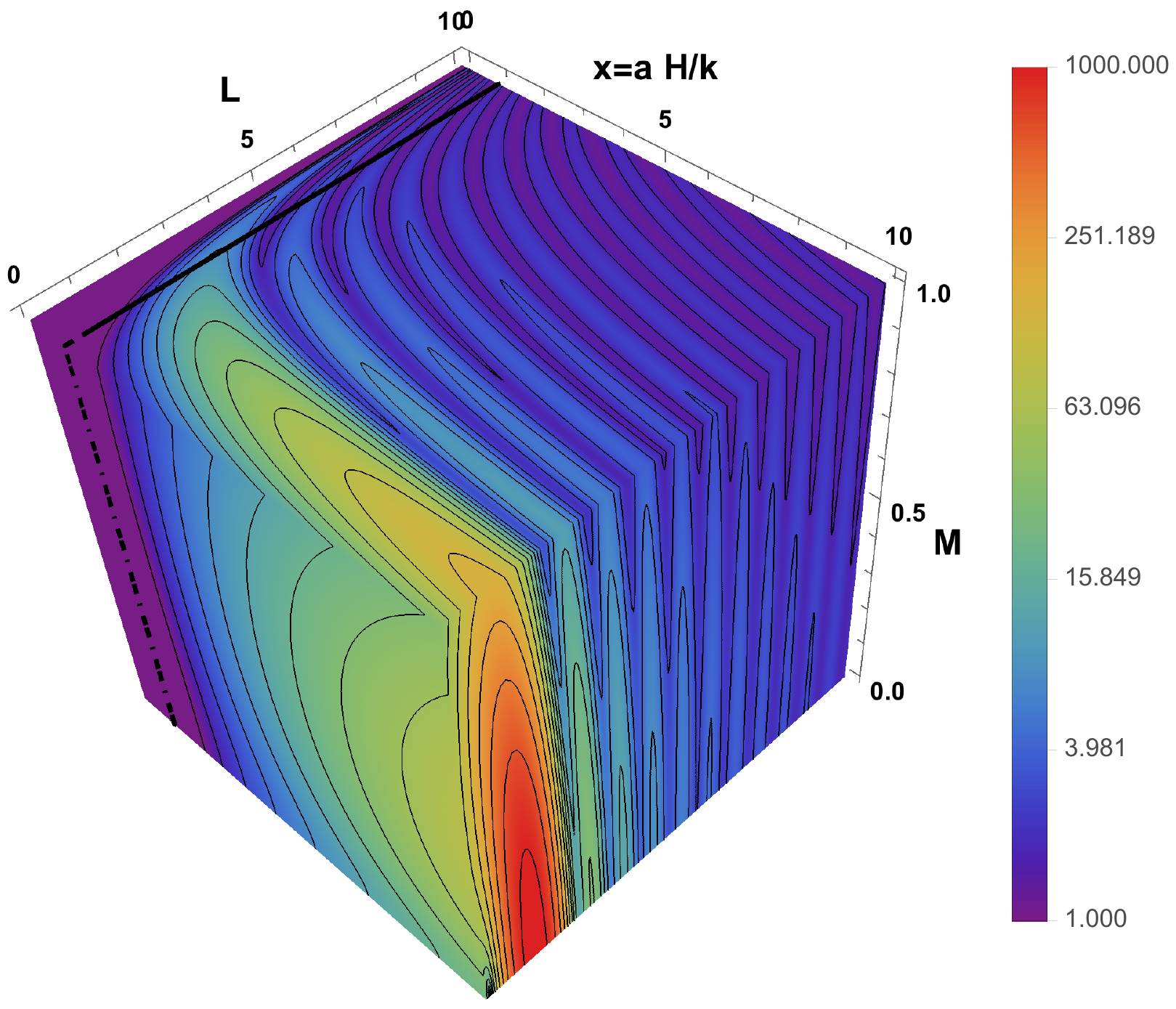}\hfill
\includegraphics[width=0.48\textwidth]{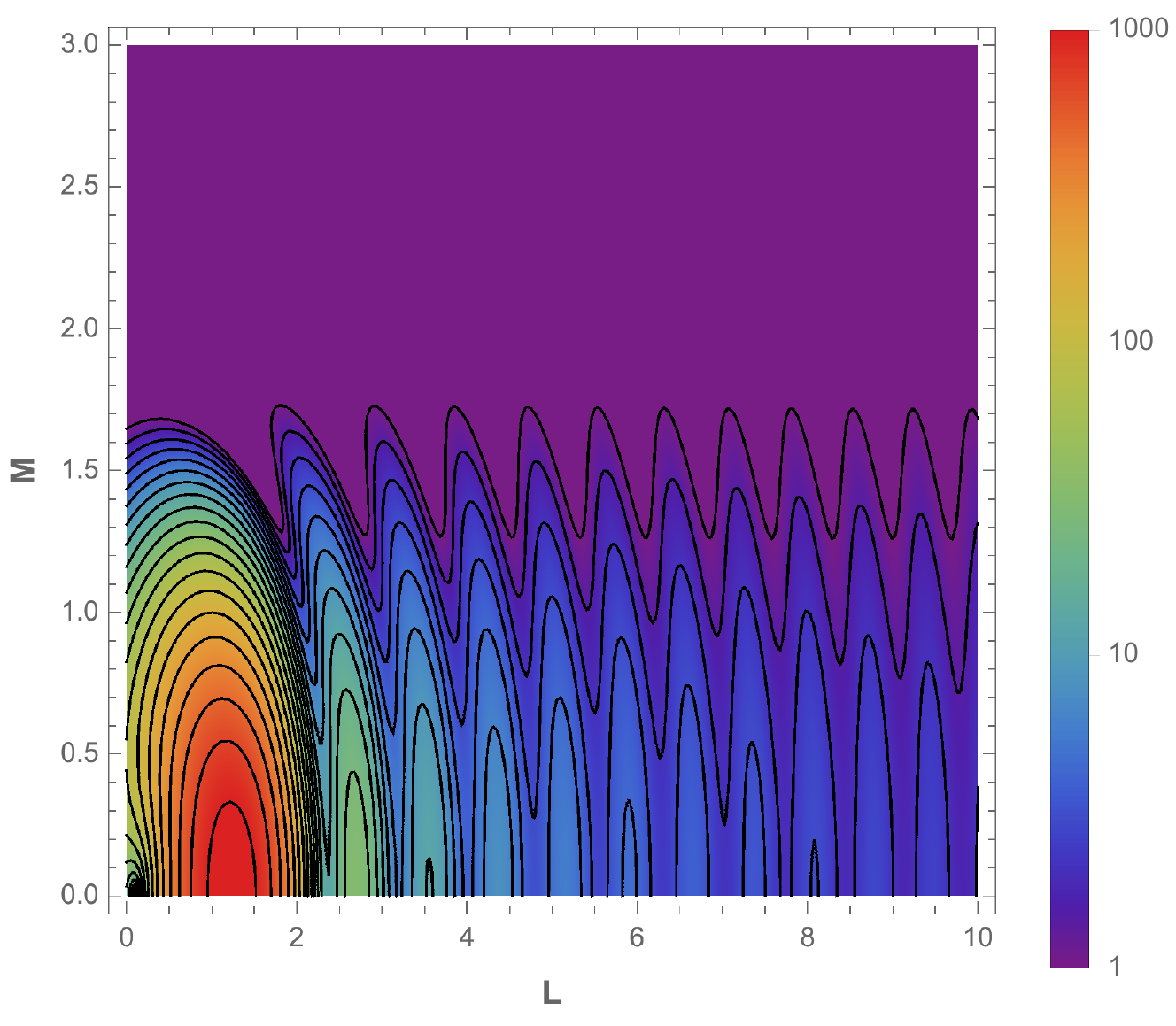}\\
\includegraphics[width=0.48\textwidth]{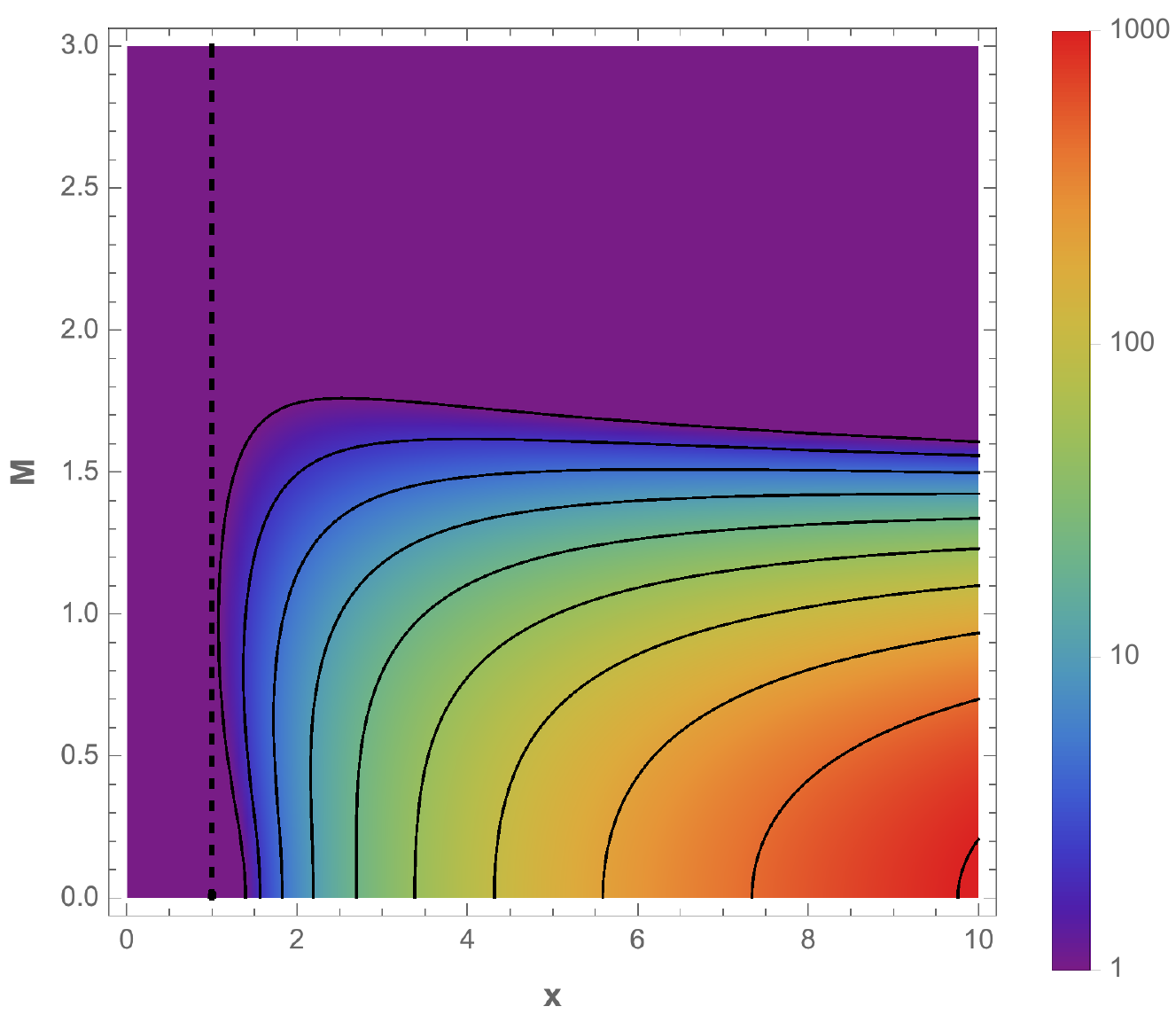}\hfill
\includegraphics[width=0.48\textwidth]{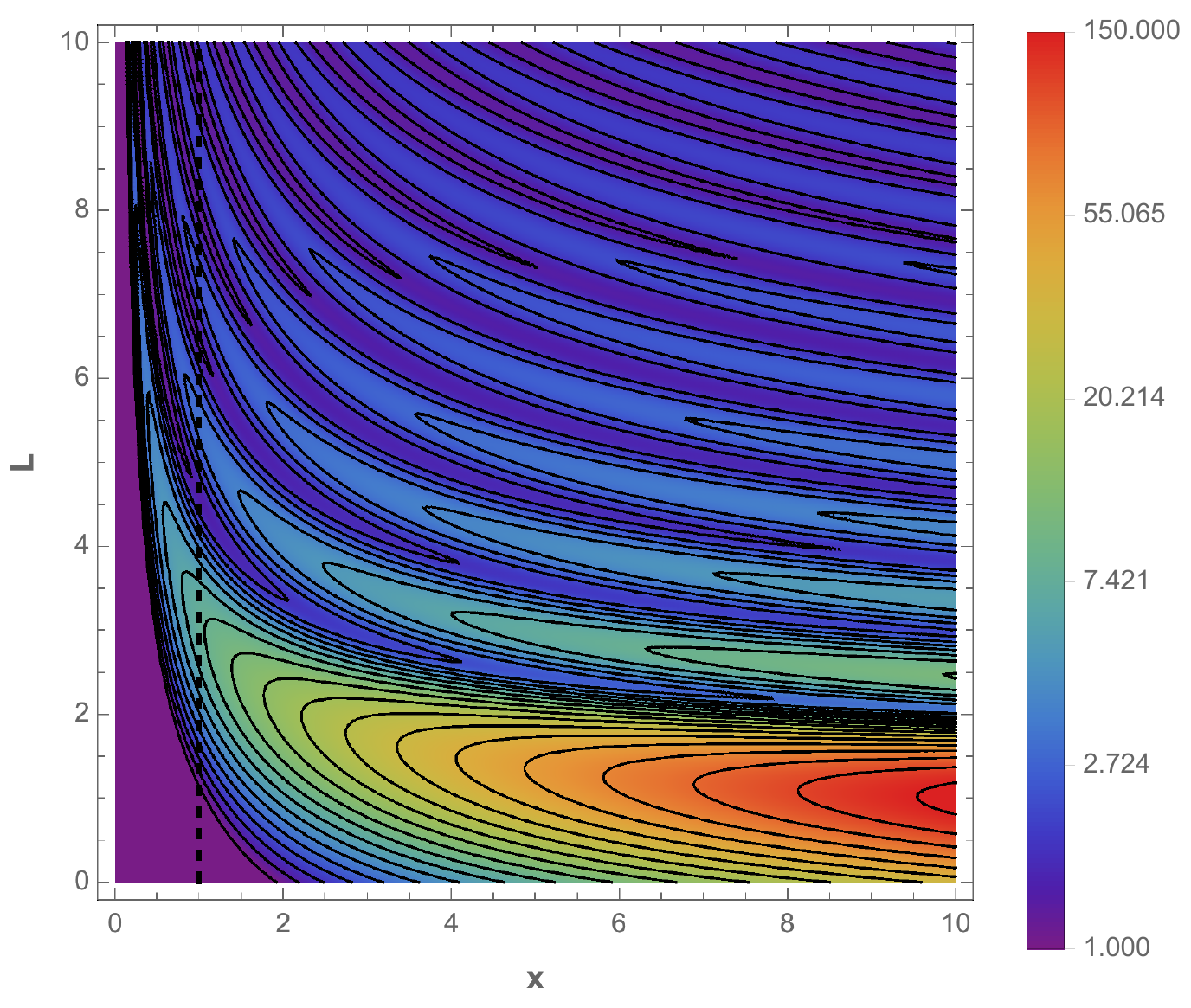}
\caption{{\bf Facets of Schwinger effect in de Sitter space.} The color shading of the Schwinger-de Sitter cube gives mean particle number, $\bra n_\tbfk \ket$ as a function of the electric field strength, $L$, mass of the field $M$, and $x= a H/k$ with $x=1$ (dashed line) is where the modes exit the comoving Hubble radius. We have $L=1$ for the $M$-$x$ plot while the rest are according to the faces of the cube. We see the particle creation is enhanced for $x>1$, weak electric field and low mass of the scalar field.}
\label{fig:faces}
\end{figure*}

In this regard, the literature cited above takes the following route to infer the particle content: one starts out with a vacuum state \cite{bunchdavies} in the asymptotic past, looks at its quantum evolution in the Heisenberg picture with canonical quantization and computes of the Bogoliubov coefficients to determine particle excitations at late times. The non-zero coefficient $\beta_\tbfk$ relating the modes in the past and the future gives the particle content and the production rate. However, this interpretation \emph{is limited to} and \emph{works only} in the ``adiabatic" regime specified here by $|eE|,m^2 \gg H^2$ (strong electric field and/or heavy fields) where the adiabatic \emph{out} vacuum can be defined. Although considered due to technical reasons, in our view, this is a highly restrictive analysis. For one, there is a wide non-adiabatic domain where the effect can be much more significant and secondly, the above discourse only provides the late-time particle content and gives no information about its evolution or its value at any \emph{given} time. The Bogoliubov coefficients can still be calculated in a time-dependent way. We, however, adopt a different path described below which brings out the effect in full colors (\fig{fig:faces}) and offers an additional advantage for studying the quantum-to-classical transition in the phenomenon as well.

\section{Schr\"odinger Formalism} 
\label{sec:formalism}

Instead of doing a canonical quantization, we proceed to perform a Schr\"odinger quantization of the system which we shall describe here briefly (see refs. \cite{gaurangA, gaurangB} for details). This formalism has been employed for studying Schwinger effect in flat spacetime \cite{gaurangB, flat} as well as particle creation in cosmological spacetimes \cite{gaurangB, suprit}. To begin with, we note that, due to translation invariance (present in most cases) the Fourier modes of the scalar field decouple to a set of harmonic oscillators:
\be 
S = \f{1}{2}\int d^3k\,dt\,  m_\tbfk(t) \left(\dot{q_\tbfk}^2-\omega_\tbfk^2 (t) q_\tbfk^2\right)
\ee with both time-dependent mass and frequency that can be treated independently. This reduces the field-theoretic problem to one in the point-particle quantum mechanics domain. To infer the quantum evolution of the system, we just need to solve the time-dependent Sch\"odinger equation associated with the oscillators: 
\be
\l[-(1/2m_\tbfk)\pp_{q_\tbfk}^2+ m_\tbfk \omega_\tbfk^2 q_\tbfk^2/2 - i\, \pp_t\r]\psi_\tbfk(q_\tbfk,t) = 0.
\label{seq}
\ee
\begin{figure*}[ht!]
\centering
\subfloat[Case: $L, M<1$]{\label{fig:ep1}
  \includegraphics[width=.45\textwidth]{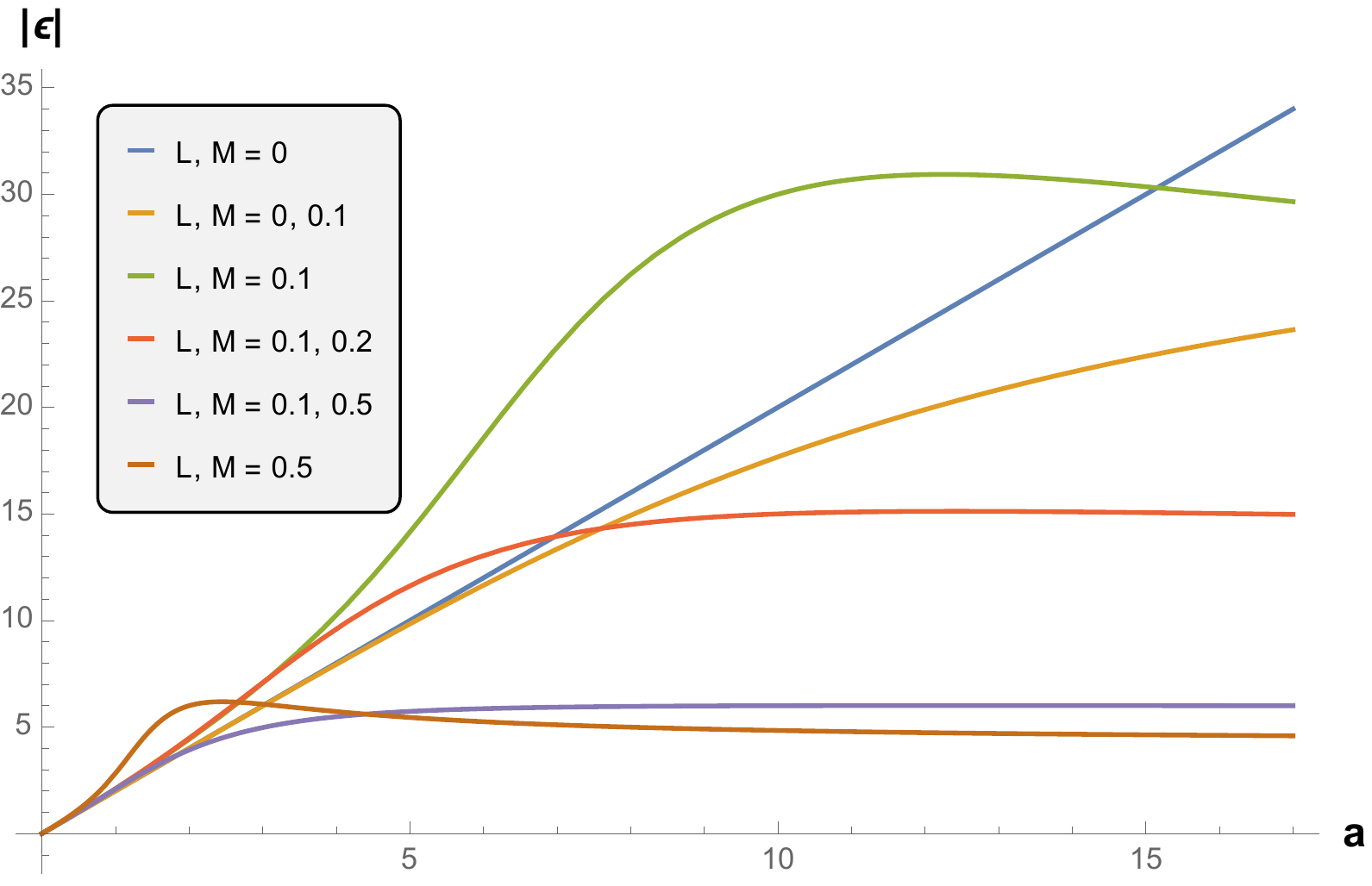}%
}~
\subfloat[Case: $L<1, M>1$]{ \label{fig:ep2}
 \raisebox{0mm}{\includegraphics[width=.45\textwidth]{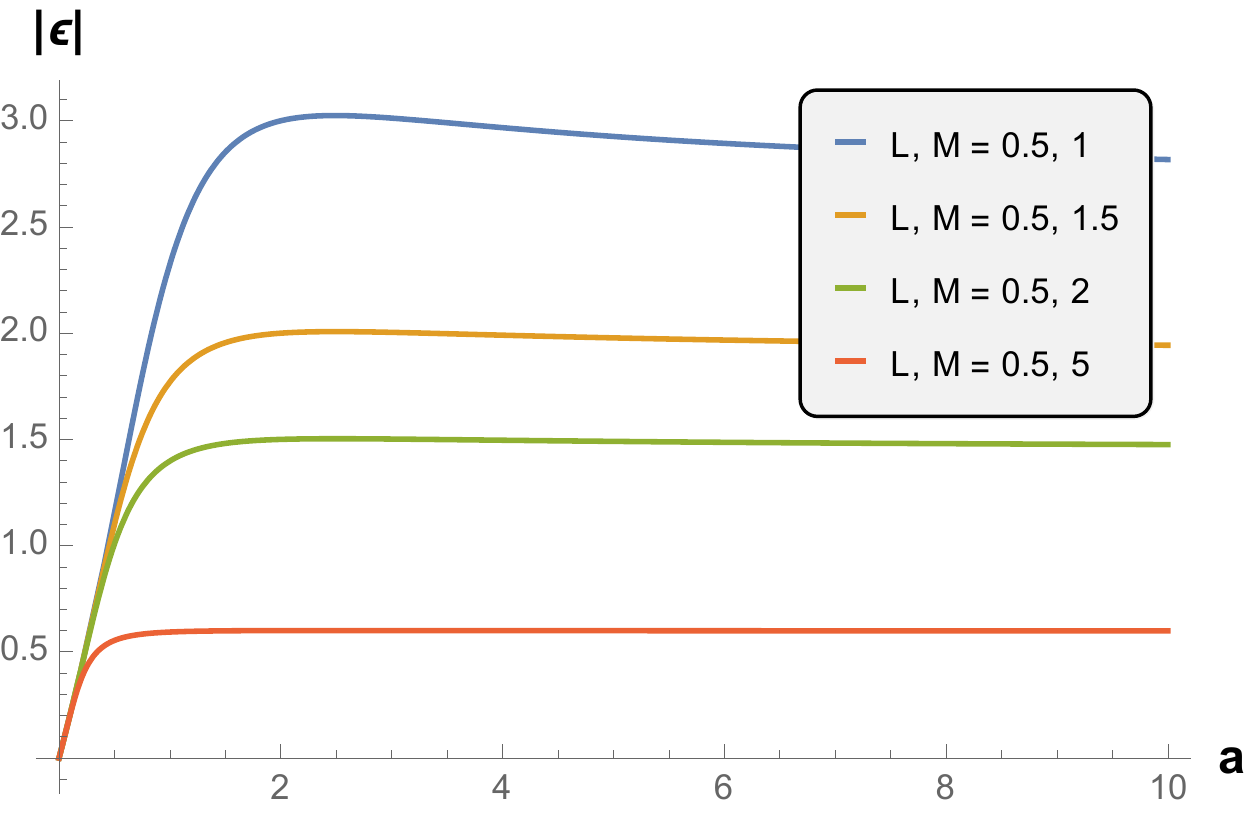}}%
}\\
\subfloat[Case: $L>1, M<1$ ]{\label{fig:ep3}
 \raisebox{0mm}{\includegraphics[width=.45\textwidth]{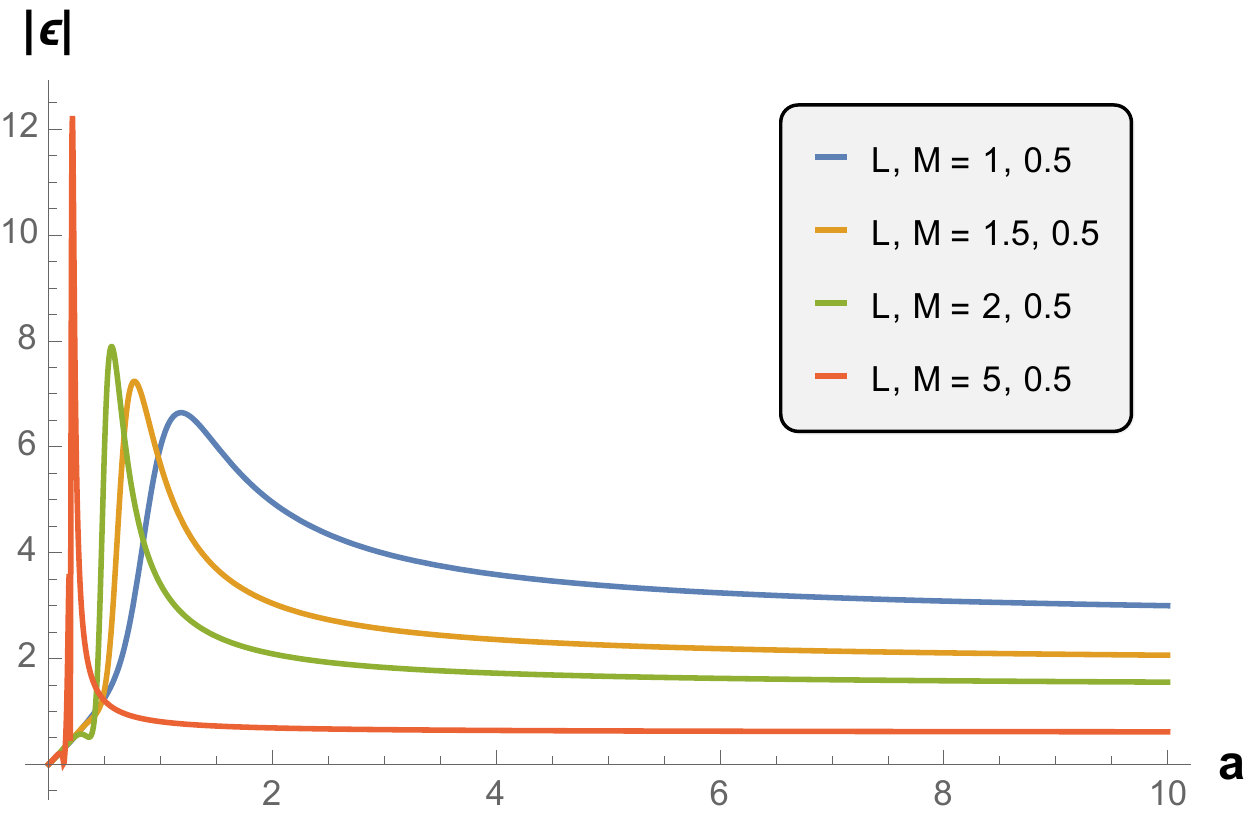}}%
}~
\subfloat[Case: $L, M > 1$]{\label{fig:ep4}
 \raisebox{0mm}{\includegraphics[width=.45\textwidth]{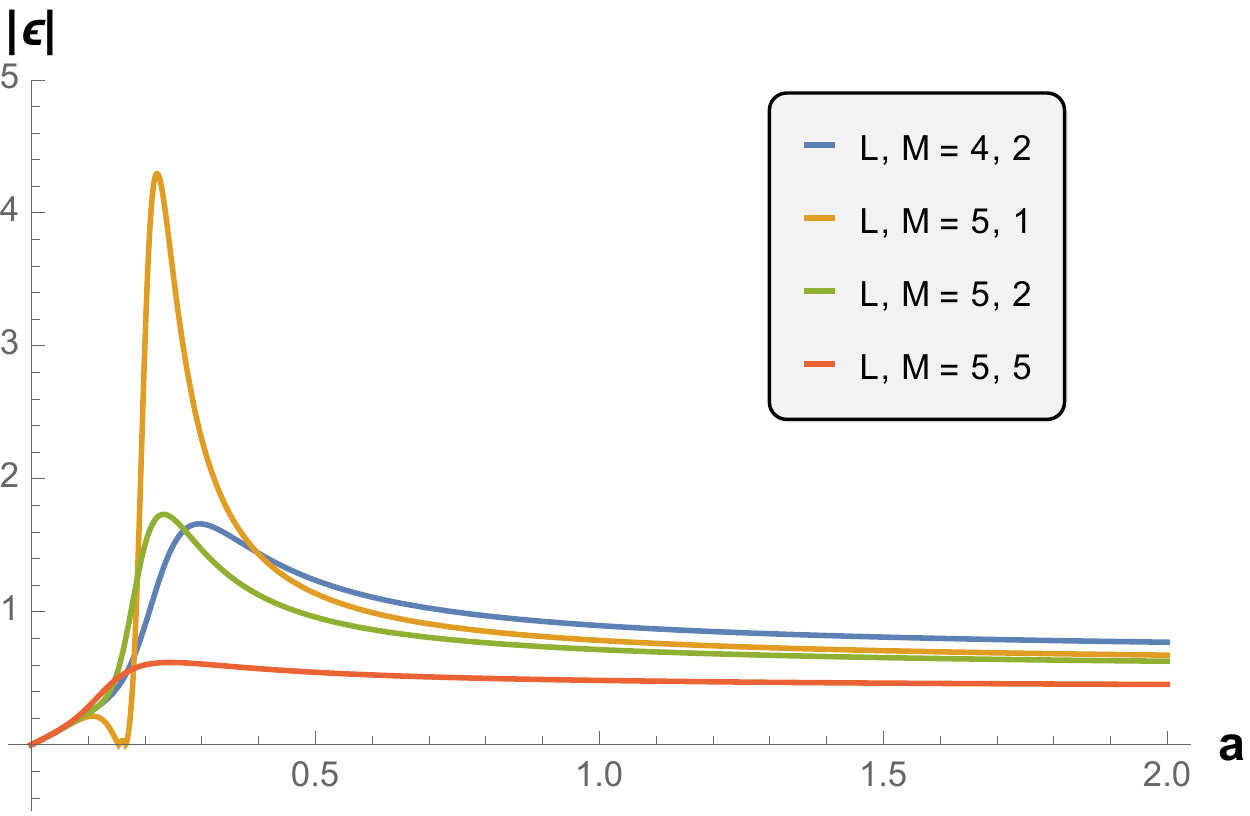}}%
}
\caption{Evolution of adiabaticity parameter $\epsilon$ with the scale factor for different $L$ and $M$ specifying field strength and mass, taking $k=k_z=1$. The system can have non-adiabatic evolution as in plots (a), (b) and (c) where the in-out prescription fails.}
\label{fig:epsilon}
\end{figure*}
\noindent The equation admits form invariant, Gaussian states: $\psi_\tbfk (q_\tbfk,t) =  N_\tbfk \exp[- \aa_\tbfk\, q_\tbfk^{2}]$ as it solutions. These are akin to coherent states of a harmonic oscillator constrained only by appropriate boundary conditions. Except for an overall phase of $N_\tbfk$, all non-trivial aspects of evolution are solely encoded in the variable $\aa_\tbfk$. We can obtain the dynamical equation for the same: 
\be
i\,\dot{\aa}_\tbfk = 2 \aa_\tbfk^2/m_\tbfk - m_\tbfk\oo_\tbfk^2/2
\ee
by substituting the Gaussian \emph{ansatz} in \eq{seq}. We introduce two new (more intuitive) variables by defining 
\begin{align}
\aa_\tbfk \equiv i m_\tbfk (\dot{\mu_\tbfk}/2\mu_\tbfk) = (m_\tbfk\oo_\tbfk/2)[(1-z_\tbfk)/(1+z_\tbfk)]\nonumber
\end{align}
The variable $z_\tbfk$ when non-zero measures the departure of $\aa_\tbfk$ from its adiabatic value. It is also related to the particle content of the state and hence termed as the excitation parameter. The evolution of $z_\tbfk$ is given by a first-order, non-linear differential equation: 
 \be 
 \dot z_\tbfk + 2 i \oo_\tbfk z_\tbfk + \oo_\tbfk \epsilon (z_\tbfk^{2} - 1)/2 = 0
 \ee
  where $\epsilon=({\dot \oo_\tbfk/\oo_\tbfk} +{\dot m_\tbfk/m_\tbfk})/\oo_\tbfk$ is the adiabaticity parameter. The dynamical equation for $\mu_\tbfk$ can also be obtained and is easier to handle analytically being a linear (though, second order) differential equation: \be 
  \ddot{\mu}_\tbfk+\l(\dot{m_\tbfk}/m_\tbfk\r)\dot{\mu}_\tbfk+\omega_\tbfk^2\,\mu_\tbfk=0
  \ee
   Remarkably, this is same as the classical equation for $q_\tbfk$ that we would have obtained by varying the oscillator action. The problem now comes down to solving these dynamical equations under appropriate initial conditions specifying the ``vacuum" at some initial time and then inferring the subsequent evolution of the wavefunction. Being all related, we can solve for any one of the variables and get the others from that. By virtue of the initial condition, our state begins as a ground state with zero particle content but at any later time it will be different from the instantaneous ground state. The \emph{instantaneous} particle content of our state, over the course of its evolution, is determined by considering its overlap with the adiabatically evolved instantaneous eigenstates defined \emph{at each} moment. On computing this overlap one finds that the mean particle number is 
\be 
\bra n_\tbfk\ket =  |z_\tbfk|^2/(1-|z_\tbfk|^2)
\ee
    and the mean value of energy at any time given by the expectation value of the Hamiltonian is 
    \be 
    E_\tbfk(t) =\big(\langle n_\tbfk \rangle +1/2\big)\,\oo_\tbfk (t)\nonumber
    \ee
     It is important to note, however, that this time-dependent mean particle number may not be monotonic in general and can have oscillatory behaviour in certain regimes. In that case, this construct should not be taken as ``particle" number in the classical sense. The system or, more so, the state that we have constructed is away from classicality and is accompanied by a certain \emph{quantum} noise. It, therefore, needs to be tied up with some measure quantifying the degree of classicality of the state (or system) for a proper interpretation. The quantity to do that -- termed the \emph{classicality parameter} -- can be constructed using the parameters of the Wigner function for our Gaussian state which is given by:
\be
W(q_\tbfk,p_\tbfk)=(1/\pi)\exp[-q_\tbfk^2/\sigma_\tbfk^2 
-\sigma_\tbfk^2 \l(p_\tbfk-{\cal J_\tbfk}\,q_\tbfk\r)^2]\nn
\ee
where $\sigma_\tbfk^2 = |1+z_\tbfk|^2/[m_\tbfk \,\omega_\tbfk\, (1-|z_\tbfk|^2)]$ and ${\cal J_\tbfk} = [2 \,m_\tbfk \,\oo_\tbfk \,{\rm Im}(z_\tbfk)]/|1+z_\tbfk|^2$. The classicality parameter is then defined and computed to be:
\be
C_\tbfk \equiv \frac{\bra p_\tbfk q_\tbfk \ket_{W}}{\sqrt{\bra p_\tbfk^2\ket_{W}\bra q_\tbfk^2 \ket_{W}}} =  \frac{{\cal J_\tbfk} \sigma_\tbfk^2}{\sqrt{1 + ({\cal J_\tbfk} \sigma_\tbfk^2 )^{2}}}
\ee
where ${\cal J_\tbfk} \sigma_\tbfk^{2} = 2 \bra p_\tbfk q_\tbfk \ket_{W}$ specifies the phase space correlation. By construction, we have $C_\tbfk \in [-1,1]$ which vanishes for a pure quantum system such that the Wigner function is an uncorrelated product of Gaussians and $|C_\tbfk| = 1$ for a classical system with a high degree of correlation in phase space. This construction is empirical and has been shown \cite{gaurangA, gaurangB, suprit} to work well for a number of cases in tight correlation with the behaviour of mean particle number defined above but is not without its limitations. This concludes our review of the formalism and the differences of our approach from the previous studies which allow us to discuss issues like time-dependent particle content and emergence of classicality \emph{without any restrictions}.

\section{A Schwinger \& de Sitter collaboration} 

We consider a minimally coupled, massive charged scalar field in the presence of a uniform electric field in spatially flat, expanding de sitter spacetime. The line element is
\begin{align}
ds^2 = a^2(\eta)(-d\eta^2+d{\bf x}^2)
\end{align}
with $a(\eta) = -1/(H\eta)$, (constant) Hubble parameter $H$ and conformal time $-\infty$ $<$ $\eta$ $<$0. The dynamics is dictated by the action: 
\be 
S=-(1/2)\int d^4x\,\sqrt{-g}\,[(D^\mu\phi)^*D_\mu\phi +m^2|\phi|^2]
\ee
where $D_\mu \equiv \partial_\mu+i e A_\mu$. Due to translation invariance of the background, the scalar field decouples to a set of harmonic oscillators, 
\be 
S = (1/2)\int d^3k\,d\eta\,  a^2(\eta) \left(|\dot{q_\tbfk}|^2-\omega_\tbfk^2|q_\tbfk|^2\right)
\ee
with time-dependent mass $a^2(\eta)$ and frequency $\omega_\tbfk^2 = k^2+e^2 A_z^2 +2ek_z A_z +m^2 a^2$. We work in the gauge where $A_\mu=(0,0,0,A_z)$ so that a constant electric field in the $z$-direction defined by $F^2= -2E^2$ gives $F_{0z} = E\, a^2$ and hence $A_z= - E/H^2\eta$. We shall solve for the variable $\mu_\tbfk$ introduced in the formalism to infer the quantum evolution. Introducing a rescaling $\tilde{\mu}_k\equiv a \mu_k$ and defining a set of new variables:
\begin{align} 
&\tau\equiv 2 i k \eta,~\kappa\equiv -i k_z L/k,~L=eE/H^2,\nn\\ 
&M\equiv m/H, ~\nu \equiv \sqrt{9/4-L^2-M^2}\nonumber
\end{align}
transforms the dynamical equation of $\mu_\tbfk$ to a well known form:
\be  
\tilde{\mu}''_\tbfk+\l\{(1/4-\nu^2)/\tt^2+\kk/\tt-1/4\r\}\tilde{\mu}_\tbfk=0
\ee
(Here $x'\equiv dx/d\tt$) which has Whittaker function $W_{\kappa ,\nu}(\tau)$ and its complex conjugate as the solutions. We now need to set the vacuum initial condition to determine the correct solution. A handle on this and the nature of subsequent evolution is provided by the adiabaticity parameter:
\begin{align}
\epsilon = \l(\left(a^3 H^3 \left(L^2+M^2\right)- a^2 H^2 k_z L\right)+a H \oo_\tbfk ^2\r)/\oo_\tbfk^3.\nn
\end{align}
\begin{figure*}[th!]
\centering
\includegraphics[scale=1]{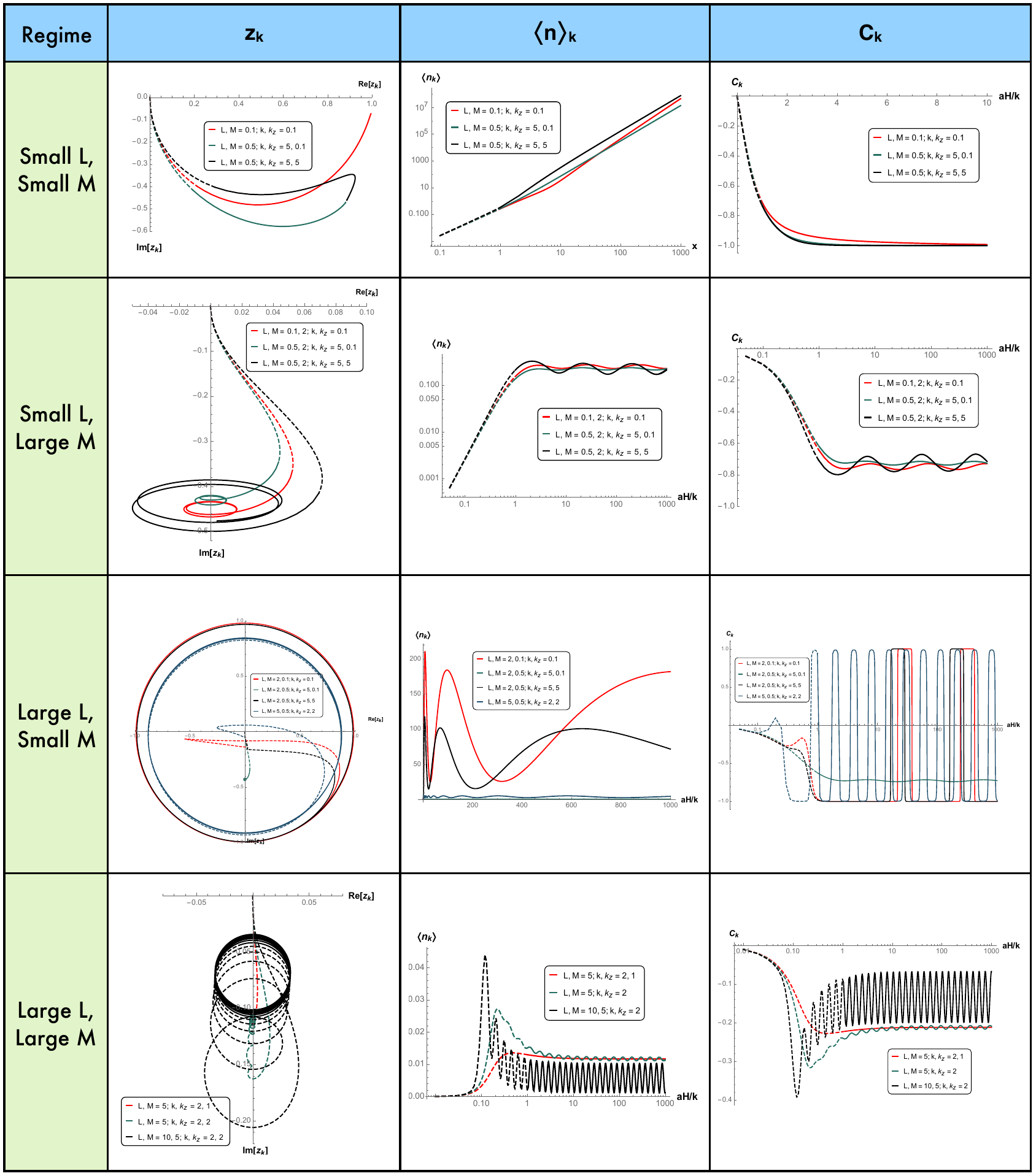}
\caption{Evolution of the excitation parameter, $z_k$, the mean number density, $\bra n_k\ket$, and the classicality parameter, $C_k$. The evolution is with respect to $x = aH/k$ such that $x<1$ is sub-Hubble (dotted lines) period and $x=1$ specifies point at the which modes exit the comoving Hubble radius to become super-Hubble for $x>1$ (regular lines).}
\label{fig:tableznc}
\end{figure*}
In the asymptotic past ($a$ $\to$0 limit), we have $\oo_\tbfk^2\simeq k^2$ and $\epsilon$ $\to$0 and thus we can define a vacuum state in that limit. This is equivalent to the condition, $\l.\dot{\mu}_\tbfk/\mu_\tbfk\r|_{\eta\to-\infty} \simeq i \omega_\tbfk$ which fixes $\tilde{\mu}_\tbfk$ to be: 
\be \tilde{\mu}_\tbfk (\tau)= (e^{i \pi  \kappa /2} /\sqrt{2 k})  \,\, W_{\kappa ,\nu }^*(\tau)
\ee
 akin to the Bunch-Davies vacuum condition. The adiabatic behaviour at late times is not guaranteed here unlike in the case of flat spacetime where $\epsilon^{\rm flat} (\tt) = \tt/[(k_\bot^2+m^2)/(e \,E)+\tt^2]^{3/2}$ vanishes for $\tt$$\to$$\pm\infty$ so that \emph{in} and \emph{out} vacuum states can be defined at early and late times. The late-time adiabatic regime exists only when $(L^2+M^2)$$\gg$1, otherwise the evolution is non-adiabatic for various cases as depicted in \fig{fig:epsilon}. We also see certain non-trivial features of high, intermediate non-adiabaticity shown in \fig{fig:ep4}. This occurs whenever $\oo_\tbfk$$\ll$1 and can also be divergent when $\oo_\tbfk=0$ which being quadratic in $a$ can be solved to get
\be
a_{\pm} = k_z L\pm\sqrt{(k_z^2 - k^2) L^2- k^2 M^2}/[(L^2+ M^2)H].\nn
\ee     
For $k_z=k$, this gives $a_{\pm} = k/[H (L\pm i M)]$ which is real only in the massless case. This also affects the particle content shows a spike at this value of the scale factor in the massless limit. It is evident from the adiabaticity parameter that Schwinger effect in de Sitter background is a rich non-adiabatic domain for weak electric field and light fields while the adiabatic vacuum exists at late times only for massive field and strong electric field background. The usual in-out prescription to obtain particle production rate fails in the case of non-adiabatic evolution. This is particularly the main emphasis of this work: to obtain a meaningful description of the (particle creation) dynamics when the quantum scalar field sub-system is away from an adiabatic evolution. 

\noindent Using $\mu_\tbfk$, we can compute all the required quantities. For instance, the excitation parameter $z_\tbfk$ is given by,   
\begin{align}
z_\tbfk = &\{[k(\oo_\tbfk+k)-a H (k_z L+i k)] W_{-\kk,\nu}(2 i k/a H)\nn\\
&+i a H k\, W_{1-\kk,\nu}(2 i k/a H)\}/\{[k (\oo_\tbfk-k)+a H (k_z L\nn\\
& +i k)]W_{-\kk,\nu}(2 i k/a H)-i a H k \,W_{1-\kk,\nu}(2 i k/a H)\}.\nn
\end{align}
This is, however, quite a complicated expression and does not give much information as to how the evolution is progressing. The exact expressions for the particle content and the classicality parameter are even more complicated and incomprehensible in their complete generality. We will resort to understand the evolution of these quantities through numerical plots and explain the features analytically in tractable limits. The results are tabulated in \fig{fig:tableznc} which has plots showing the evolution of $z_\tbfk$, $\bra n_\tbfk\ket$ and $C_\tbfk$. The dashed and normal lines correspond to the sub-Hubble ($aH/k<1$) and super-Hubble ($aH/k>1$) phases respectively. Also, note that the evolution in the plots is shown with respect to $x = aH/k$ rather than the scale factor. 

{\bf Regime I: $\boldsymbol{L,M\ll1}$ \fig{fig:tableznc} (first row)}. We have a weak external electric field and light scalar field in this case. The evolution of $z_\tbfk$ in its complex plane is similar to its evolution of massless scalar field in pure de Sitter spacetime \cite{suprit}. For a weak field, the evolution at late-times is highly non-adiabatic with $z_\tbfk$ close to unity that shows up in the particle content which increases monotonically. At late-times $\bra n_\tbfk\ket \propto (a H/k)^{2\nu}$ with $\nu$ real and finite giving straight lines in the logarithmic plot. The classicality parameter starts from zero and grows to $(-1)$ as the modes exit the Hubble radius ($aH/k>1$), that is, in concordance with the emergence of classicality on Hubble exit. Further, the differences in the plots due to different $k$ and $k_z$ values show up only in the super-Hubble phase. 

{\bf Regime II: $\boldsymbol{L<1< M}$ \fig{fig:tableznc} (second row)}. In this case, we have a weak, external electric field but heavy scalar field. The evolution of $z_\tbfk$ in its complex plane starts off from the origin but gets subsequently locked on the imaginary axes going around in circles. The regime is mildly non-adiabatic and becomes adiabatic for large $M$. The particle content is suppressed greatly and saturates in the super-Hubble phase with slight oscillations. The classicality parameter also shows these slight oscillations on Hubble exit and saturates, but, below the complete classical limit of maximum correlation. 

{\bf Regime III: $\boldsymbol{M<1< L}$ \fig{fig:tableznc} (third row)}. With strong field and low mass limit, the evolution actually resembles that in the case of flat spacetime background \cite{gaurangB}. The evolution of $z_\tbfk$ is close to that of a rotor in the complex plane, that is, it circles around with a near-constant magnitude. The particle number is negligible in the sub-Hubble phase and then grows up sharply as the modes make an exit, oscillates and saturates at late times. The classicality parameter grows to minus one and then oscillates between its extremities and does not give a clear idea of the degree of classicality in this case, although it should be noted that the variance of $C_\tbfk$ is finite and non-zero.   

{\bf Regime IV: $\boldsymbol{L,M \gg1}$ \fig{fig:tableznc} (fourth row)}. With large $L$ and $M$, we can define the adiabatic vacuum at late times since the adiabaticity parameter $\epsilon<1$. Also the parameter $\nu$ is purely imaginary in this case so that we can write it as $\nu = i\, |\nu|$, that is, the arg $\nu =\pi/2$. The parameter $z_\tbfk$ is concentrated on the imaginary axis with real part being close to negligible. The particle content in this adiabatic regime can be computed exactly in the late-time limit which turns out to be:
\be
\bra n_\tbfk\ket=\f{e^{2\pi |\kk|}+e^{-2\pi |\nu|}}{2|\nu| \sinh 2\pi |\nu|} + A e^{\pi |\kk|}(9\cos\xi-6|\nu|\sin \xi)\nn\label{nk}
\ee
where $\xi = 2\theta-\phi+\psi+\chi$, $\phi \equiv \arg \Gamma(1/2-i|\kk|-i|\nu|)$, $\theta \equiv \arg \Gamma(-2i|\nu|)$, $\psi \equiv \arg \Gamma(1/2+i|\kk|-i|\nu|)$, $\chi \equiv 2 |\nu|\log(2/x)$ and
$A=[\cosh \pi (|\kk|+|\nu|)\cosh \pi (|\kk|-|\nu|)]^{1/2}/(4 |\nu|^2 \sinh2|\nu|)$.
The first term matches the result obtained in ref. \cite{Kobayashi:2014} in the adiabatic limit. The second term, which is the reason for the oscillations seen in \fig{fig:tableznc}, is however absent in their analysis. The oscillations are a result of using the adiabatically evolved instantaneous eigenstates to compute the particle content. The classicality parameter also presents a contrary feature. It shows an increase initially going towards it extreme value but then decreases and oscillates at a lower value. While the system is still away from a ``pure" quantum depiction in its later stages, this type of behaviour was not anticipated and is quite intriguing. So either the classicality parameter is amiss and insufficient to provide the correct picture or we have something else interesting going on. This requires a further analysis with some other constructs that specify the quantum to classical transition such as the quantum discord \cite{discord} etc. 

\section{Summary} We studied Schwinger effect in de Sitter space in an analysis that was expansive and unlike anything that has been carried out before. We used the Schr\"odinger quantization formalism that allowed us to go beyond the adiabatic regime and explore the effect in its full generality. Figs. \ref{fig:faces} and \ref{fig:tableznc} form the main results of this paper showing the evolution of state, its instantaneous particle content, and the degree of classicality through the classicality parameter for different cases. The deviation from the adiabaticity gives rise to particle creation which is profound in the case of \emph{weak electric field} and light scalar fields. This is contrary to expectation that particle production will be larger for a strong electric field background. The mean particle number for a given $k$-mode shows a monotonic increase only in the non-adiabatic case and is suppressed in the other regimes. The non-adiabatic domain of Schwinger effect can, thus, have significant consequences in the generation mechanisms of primordial magnetic fields. The classicality parameter shows a tight correlation with the mean particle number and exhibits an oscillatory character when the latter does too. In the non-adiabatic regime, with significantly high particle number, the system reaches a classical description specified by the classicality parameter going to $-1$ as the modes exit the Hubble radius. In the other regimes, the ``emergence" of classicality for a mode does seem to occur but the situation is not completely clear due to large oscillations at times. This points at the need of a further analysis including a look at some other methods to study the quantum to classical transition. Finally, we note that the formalism and the subsequent trail of our study can also find potential applications in the domain of analogue gravity where the experimental verifications of quantum effects are underway \cite{analogue}.

\begin{acknowledgements}
Research of R.S. is supported by SRF grant from CSIR (India) and that of S.S. by SERB OPDF (India). We acknowledge IRC, Delhi and thank Prof. T. R. Seshadri and Sunil Malik for discussions and comments on the draft.
\end{acknowledgements}

\end{document}